\documentclass{ws-ijmpe}

\def \pt {p_\perp}
\def \snn {\sqrt{s_{_{NN}}}}
\def \dphi {\Delta\phi}
\def \deta {\Delta\eta}

\begin{document}

\markboth{Fuqiang Wang}{Forward- and Mid-Rapidity Jet-like correlations}

\catchline{}{}{}{}{}

\title{Forward- and Mid-Rapidity Jet-like correlations}

\author{\footnotesize Fuqiang Wang\\
(for the STAR Collaboration)}

\address{Department of Physics, Purdue University\\
525 Northwestern Avenue, West Lafayette, Indiana 47907, USA\\
fqwang@purdue.edu}

\maketitle

\begin{history}
\received{(received date)}
\revised{(revised date)}
\end{history}

\begin{abstract}
Mid-rapidity azimuthal correlations probe di-jets originating mainly from gluon-gluon hard-scattering. Measurements of such correlations have revealed significant (gluon-)jet modification in central Au+Au collisions. Azimuthal correlations of hadrons at forward rapidity with a mid-rapidity high-$\pt$ hadron, on the other hand, 
probe asymmetric partonic scatterings involving large-x quarks and small-x gluons. We present preliminary results from STAR on correlations of charged hadrons at forward rapidity in the forward TPCs ($2.7<|\eta|<3.9$, $\pt<2$~GeV/$c$) with high-$\pt$ charged hadrons at mid-rapidity from the main TPC ($|\eta|<1$, $\pt>3$~GeV/$c$) in $pp$, d+Au, and Au+Au collisions at $\snn=200$~GeV. The implications of the results for small-x gluon distributions (Color Glass Condensate formation) and the energy loss of quark jets at forward rapidity in nuclear medium are discussed.
\end{abstract}

\section{Introduction}
\label{intro}

Jet-like azimuthal correlations have shown significant modifications in central Au+Au collisions at RHIC due to the presence of the hot and dense medium created in these collisions~\cite{whitepapers}. This constitutes a strong evidence of jet quenching where high energy partons lose energy through interactions with the medium. The inferred medium energy density is sufficiently high that formation of the Quark-Gluon Plasma is plausible~\cite{whitepapers}.

Most jet-like correlation measurements performed so far are at mid-rapidity where gluon-jets dominate at RHIC energies~\cite{whitepapers}. Measurements of mid-rapidity jet-like correlation with forward high-$\pt$ hadrons have been performed for $pp$ and d+Au collisions~\cite{FPD,PHENIXmuon}. 
STAR also has the capability to carry out correlation measurements at forward rapidities using the Forward Time Projection Chambers (FTPCs), which cover the pseudo-rapidity range of $2.7<|\eta|<3.9$, with mid-rapidity trigger particles from the main TPC. These correlations should be dominated by quark jets, as discussed below. This paper presents such measurements. They potentially address two problems as follow.

\subsection{Jet Quenching at Forward Rapidity}

Mid-rapidity jet-correlation measurements have revealed a number of novel phenomena: (i) The away-side jet-like correlation is significantly broadened in central Au+Au collisions. For some kinematic regions, the away-side correlation may even peak away from $\pi$ (opposite direction to the trigger particle in azimuth)~\cite{jetspectra,otherpapers}. (ii) Recent 3-particle azimuthal correlation results from STAR reveal structures consistent with conical emission~\cite{UleryHP06}. (iii) The near-side jet-like correlation at mid-rapidity is observed to have a long range $\deta$ correlation, rather flat within the measured range of $|\deta|^{<}_{\sim}1.5$ (the so-called ridge)~\cite{Puschke}. Will these phenomena also be present at forward rapidities? How far does the ridge extend in pseudo-rapidity? These questions may be addressed by jet-like correlation measurements at forward rapidities.

Gluon energy loss in nuclear medium is predicted by Quantum Chromodynamics (QCD) to be stronger than quark's by a factor of 9/4~\cite{pQCDeloss}. To test the QCD energy loss picture, it is valuable to also have measurements where quark jets dominate, which can be obtained from measurements at forward rapidities.
If quarks indeed lose less energy than gluons, one would expect that jet-like correlations at forward rapidities are less modified than at mid-rapidity, where gluon fragmentation dominates. However, the pathlengths through the medium may also be different for jets at forward and mid-rapidity. Detailed calculations incorporating realistic medium denisties and expansion dynamics are likely needed to disentangle these effects.


\subsection{The Color Glass Condensate}

The mid-rapidity di-jet correlations measured at RHIC stem primarily from hard scatterings of partons of small Bjorken $x$~\cite{whitepapers}. They are dominated by gluons because at small $x$ gluons far outnumber quarks~\cite{PDF}. On the other hand, a di-jet with one at mid-rapidity and the other at forward rapidity comes from parton scattering of quite different kinematics. While the mid-rapidity jet is still dominated by small-$x$ gluons, the forward jet has to come from a large-$x$ parton in order to be at forward rapidity with balancing $\pt$. At large $x$, valence quarks dominate~\cite{PDF}. Thus, a mid- and forward-rapidity di-jet primarily comes from hard-scattering between a small-$x$ gluon and a large-$x$ quark. For mid-rapidity trigger particle $\pt>3$~GeV/$c$ reported here, the initial gluon energy is of the order of 5~GeV, or $x_g\sim 0.05$; that for the quark is of the order $x_q\sim x_g\sinh(\eta)\sim 0.7$, based on leading order estimate.

The asymmetric parton scattering kinematics, coupled with the asymmetric d+Au collisions, provide an opportunity to probe the Color Glass Condensate (CGC)~\cite{CGC}. 
If the relative number of small-$x$ gluons to large-$x$ quarks in Au-nucleus equals to that in deuteron, there will be equal numbers of g(d)+q(Au) and g(Au)+q(d) scatterings. Normalized by the total number of trigger particles (gluon-jets), the jet-like correlation strengths at the d-side and the Au-side will be the same and add up to the correlation strength of per-quark ``fragmentation". If, on the other hand, the number of gluons is reduced relative to that of quarks in Au-nucleus, then there will be fewer g(Au)+q(d) than g(d)+q(Au) scatterings. 
As a result, the d-side correlation strength will be reduced relative to that at the Au-side. Such a reduction would be a signature of CGC.

\section{Analysis and Systematic Uncertainties}
\label{analysis}

The two main detectors used in this analysis are the STAR's main TPC~\cite{TPC} and the FTPCs~\cite{FTPC}. The analysis selects high-$\pt$ trigger particles, $3<\pt^{\rm trig}<10$~GeV/$c$, from the main TPC within $|\eta^{\rm trig}|<1$, and correlates them with charged particles measured in the FTPCs within $2.7<|\eta|<3.9$. The $\pt$ cut for the associated particles is $0.2<\pt<2$~GeV/$c$, where the upper cut is dictated by the relatively poor momentum resolution of the FTPCs. This also prevents us from selecting high-$\pt$ trigger particles from the FTPCs.

The combinatorial background is obtained from event-mixing. The elliptic flow modulation for Au+Au collisions, $2v_2^{\rm trig}(\pt^{\rm trig})v_2(\pt)\cos(2\dphi)$, is added in pairwise during event-mixing. The trigger particle elliptic flow $v_2^{\rm trig}$ is taken to be the average from the modified reaction plane and the 4-particle cumulant methods, and the range of the two is used as an estimate of the systematic uncertainty, as in~\cite{jetspectra}. The associated particle $v_2$ is measured for 20-70\% Au+Au collisions in the FTPCs by several methods~\cite{v2MRP}. 
In this analysis the two-particle cumulant $v_2\{2\}$ is used and parameterized as $v_2(\eta,\pt) = 0.0716\,(1-\exp[-(\pt/0.50)^{1.81}]) \times 2.29\,\exp[-(\eta/2.51)^2/2]$. The centrality dependence of the FTPC $v_2$ is parameterized using preliminary STAR results. 
The FTPC $v_{2}$ results obtained from different methods are in good agreement~\cite{v2MRP}. However, as a conservative estimate the same relative systematic uncertainty as at mid-rapidity is applied on the FTPC $v_2$.

The results reported here have been corrected for tracking efficiency and acceptance of associated particles, and normalized per trigger particle. Besides $v_2$, the other major source of systematic uncertainty comes from background normalization which is described in section~\ref{results}. An additional overall systematic uncertainty of 10\% is estimated due to efficiency and acceptance corrections.


\section{Results and Discussions}
\label{results}

We first present and compare the forward-rapidity jet-like correlation results from $pp$ and d+Au collisions at $\snn=200$~GeV. The d+Au results potentially probe the CGC. We then compare these results to Au+Au collisions at $\snn=200$~GeV. The question to address by this comparison is whether or not the energy loss and the jet quenching picture differ between mid- and forward rapidities. The results have been presented previously by STAR~\cite{WangHP06}.

\subsection{Results from $pp$ and d+Au}

\begin{figure}[htbp]
\centering{
\includegraphics[width=0.75\textwidth,bbllx=0,bblly=15,bburx=565,bbury=440]{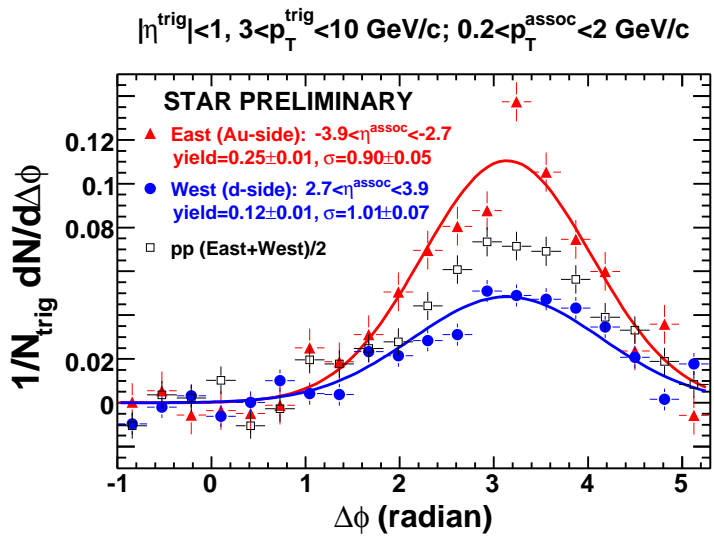}
}
\caption{(color online) Azimuthal correlations in d+Au collisions at $\snn=200$~GeV, separately for the d-side (blue circles) and the Au-side (red triangles), and that averaged over positive and negative pseudo-rapidities in $pp$ collisions (open squares). The subtracted backgrounds are obtained from event mixing and normalized to the signal in the entire near-side region of $|\dphi|<1$. The curves are Gaussian fits to the d+Au data.}
\label{fig:ppdAu}
\end{figure}

Figure~\ref{fig:ppdAu} shows the azimuthal correlations in minimum bias $pp$ and d+Au collisions. The positive (d-side) and negative (Au-side) pseudo-rapidity results are presented separately for d+Au collisions; the two $\eta$ regions are averaged for $pp$. The combinatorial background is normalized to the signal in the near-side region of $|\dphi|<1$ where no correlation signals are expected. (The near-side correlation is contained in the vicinity of the trigger particle which is restricted at mid-rapidity in the main TPC.) While the correlation shapes are the same, the correlation strength at the d-side is about a factor of 2 smaller than that at the Au-side. The $pp$ data is approximately equal to the average of the d- and Au-side correlation strengths. This is expected as the sum of the correlation strengths, to first order, should be equal to that from a single quark fragmentation as discussed above. Our measured reduction in the d-side correlation strength is in qualitative agreement with the observed high-$\pt$ suppression in particle production at forward rapidity by BRAHMS~\cite{brahms}.

It is well known that parton distributions in bound nucleons are different from the ones in free nucleons, so some reduction in the correlation strength at the d-side may be expected if the small-$x$ gluons in Au are suppressed due to shadowing. However, gluons at $x\sim 0.05$ are in the anti-shadowing region according to EKS98~\cite{EKS98} -- they are more abundant in heavy nuclei than in free nucleons by about 15\%, rather scale-independent. Moreover, the quark distribution at large $x$ is suppressed in nuclei, the so-called EMC effect~\cite{EMC}; the suppression is maximal (about 20\%) at $x\sim 0.7$. Both the gluon anti-shadowing and the EMC effect should make the relative number of small-$x$ gluons to large-$x$ quarks larger in heavy nuclei than in free nucleons, the opposite of the observed reduction of the d-side correlation strength. CGC calculcations, on the other hand, predict a suppression of the gluon density in nuclei at small-$x$~\cite{CGC}. The observed reduction of the d-side correlation strength is in qualitative agreement with such a suppression, and is also in line with the mid-rapidity correlation measurement with a forward $\pi^0$ by STAR~\cite{FPD}. However, we note that the forward quark from the deuteron may suffer energy degradation (nucleon rapidity shift) traversing the Au-nucleus thickness before striking a gluon from the Au-nucleus. This could also reduce the d-side jet-like correlation strength, whose magnitude, however, needs theoretical investigation.
\subsection{Results from Au+Au}

\begin{figure}[htbp]
\centering{
\includegraphics[width=\textwidth]{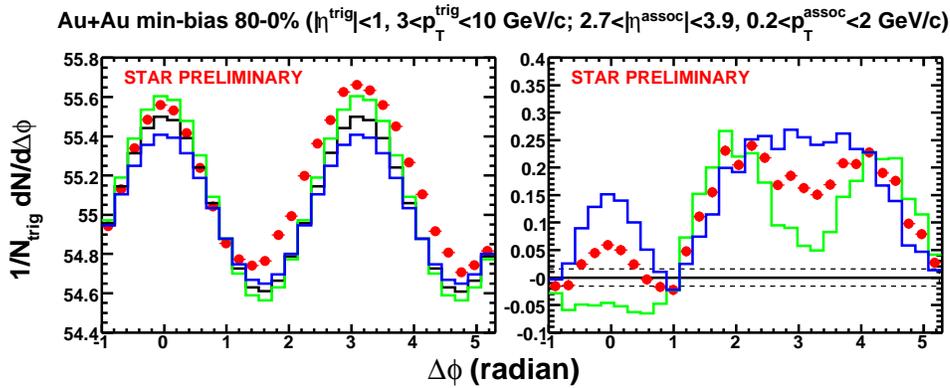}
}
\caption{(color online) Raw (left panel) and background-subtracted (right panel) azimuthal correlations in minimum bias Au+Au collisions at $\snn=200$~GeV. The histograms in the left panel are flow-modulated combinatorial background normalized to the signal within $0.8<|\Delta\phi|<1.2$ (black) and the corresponding systematic uncertainties (blue and green) due to elliptic flow. The histograms in the right panel are the systematic uncertainties on the correlation result due to elliptic flow, and the dashed lines indicate those due to background normalization.}
\label{fig:AuAuMB}
\end{figure}

Figure~\ref{fig:AuAuMB} shows the azimuthal correlation results in minimum bias Au+Au collisions (about 80\% of the total geometrical cross-section). The raw correlation is shown in the left panel together with the combinatorial background. The background-subtracted result is shown in the right panel. 
The background is normalized to the signal within the normalization range of $0.8<|\Delta\phi|<1.2$ by the Zero Yield At 1 (ZYA1) method~\cite{jetspectra,method}, where the minimum correlation signal is found. The systematic uncertainty on the background normalization is estimated by varying the size of the normalization range around $\dphi=1$. We do not normalize the background to the entire near-side region as for $pp$ and d+Au because the near-side correlation in Au+Au may not be zero {\em a priori}; the $\deta$ ridge observed at mid-rapidity~\cite{Puschke} could extend to forward rapidities. As seen from the right panel of Fig.~\ref{fig:AuAuMB}, the near-side correlation for $0.2<\pt<2$~GeV/$c$ in minimum bias data is consistent with zero within the systematic uncertainties.

\begin{figure}[htbp]
\centering{
\resizebox{.49\textwidth}{!}{\includegraphics{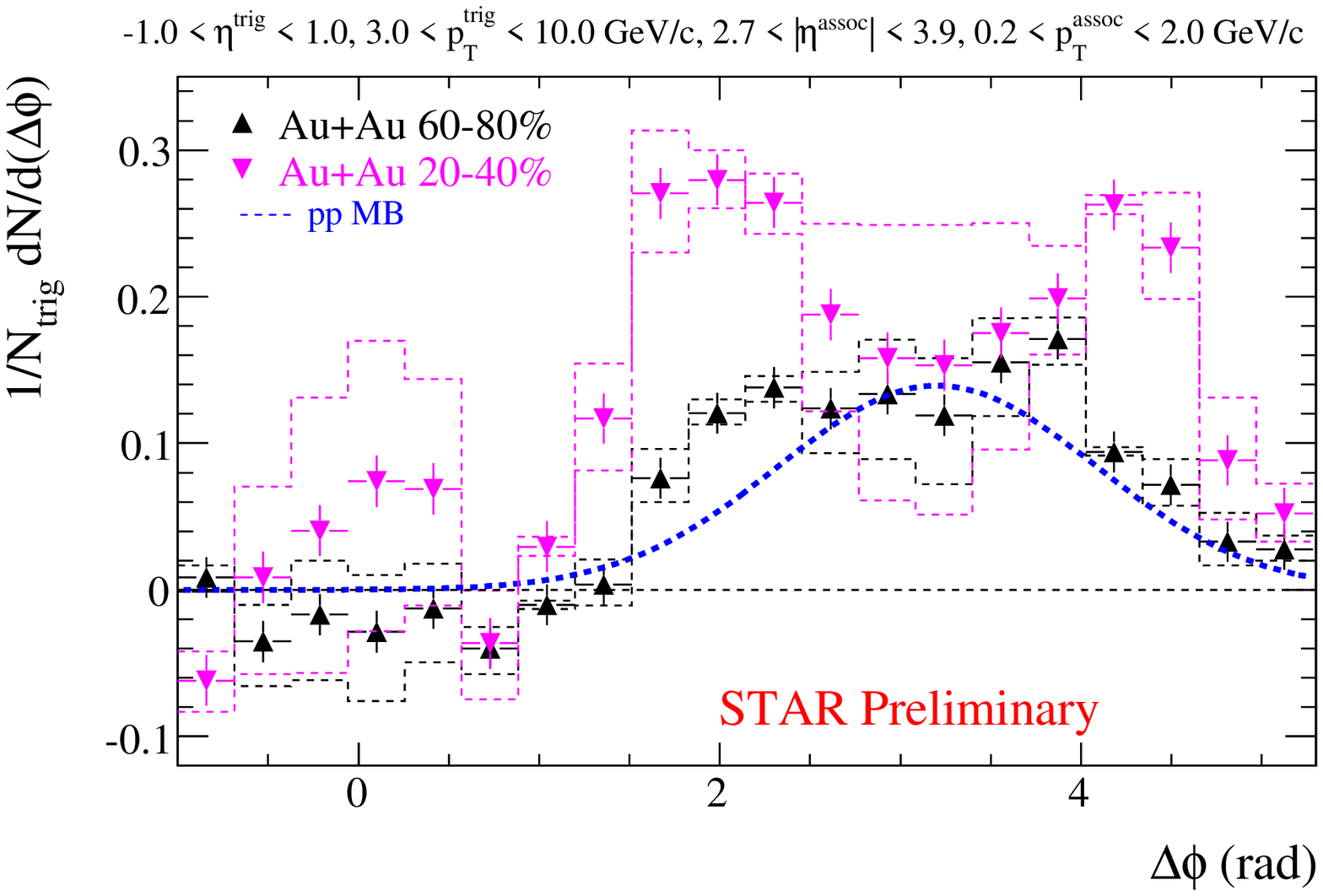}}
\resizebox{.49\textwidth}{!}{\includegraphics{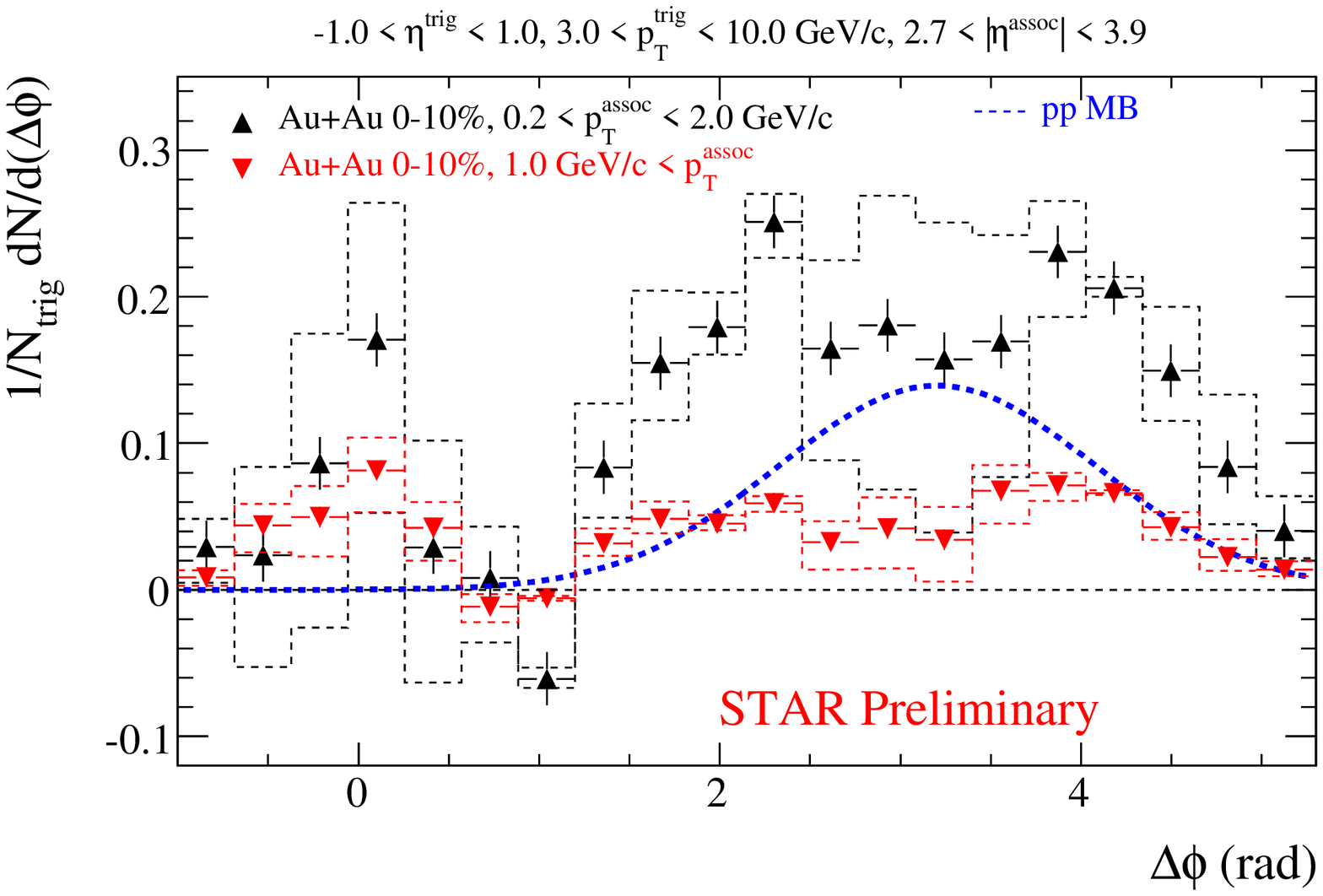}}
}
\caption{(color online) Left panel: correlation functions in 60-80\% and 20-40\% Au+Au collisions at 200~GeV/$c$. Right panel: correlation functions for two associated momentum ranges in 0-10\% Au+Au collisions at 200~GeV/$c$. The dashed smooth curve represents a fit to the correlation function in $pp$ collisions at $\snn=200$~GeV/$c$.}
\vspace*{-0.3cm}
\label{fig:AuAu}
\end{figure}

Figure~\ref{fig:AuAu} shows the azimuthal correlations for 60-80\% and 20-40\% (left panel) and for 0-10\% (right panel) Au+Au collisions. Results for two associated $\pt$ ranges are shown for the 0-10\% Au+Au data. The near-side correlations for $0.2<\pt<2$~GeV/$c$ are consistent with zero within the systematic uncertainties. However, the central data at high associated $\pt$, with the reduced systematic uncertainty, are suggestive of non-zero correlation on the near-side. This result indicates that long range $\deta$ correlations, first observed in $|\deta|^{<}_{\sim}1.5$~\cite{Puschke}, may extend out to $\deta\sim 4$ in the FTPCs.

The away-side correlation shapes in Au+Au collisions are significantly broadened relative to $pp$ and d+Au collisions as shown in Fig.~\ref{fig:AuAu}. The broadening is present for each centrality and is similar to mid-rapidity measurements. Given the present systematic uncertainties, the away-side correlation is consistent with either a flat or a double-peaked distribution. However, the observed broadness of the away-side correlation is robust and suggests that the medium modification to jet-like correlations is as strong at forward rapidity as at mid-rapidity. 

\begin{figure}[htbp]
\centering{
\resizebox{.327\textwidth}{!}{\includegraphics{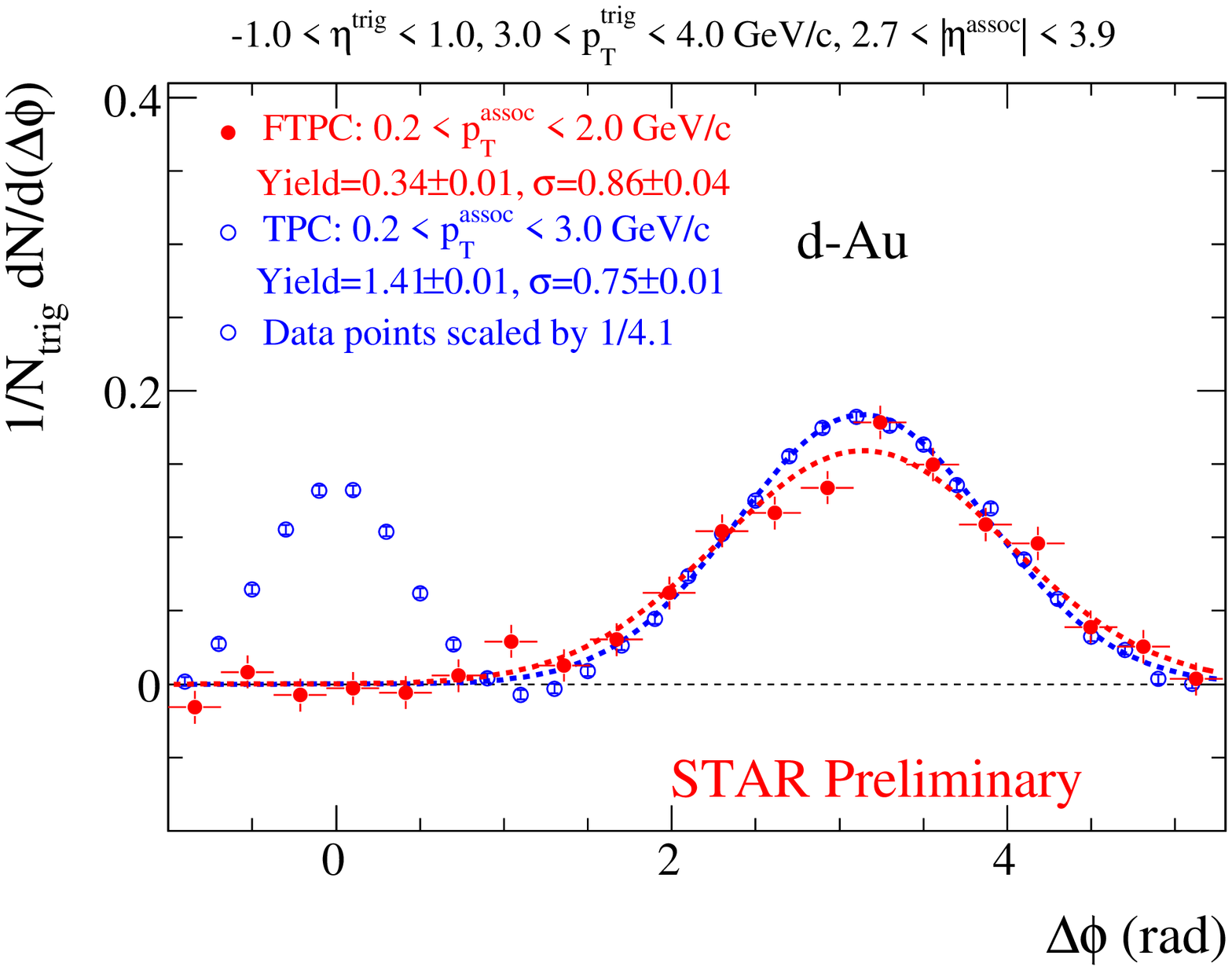}}
\resizebox{.327\textwidth}{!}{\includegraphics{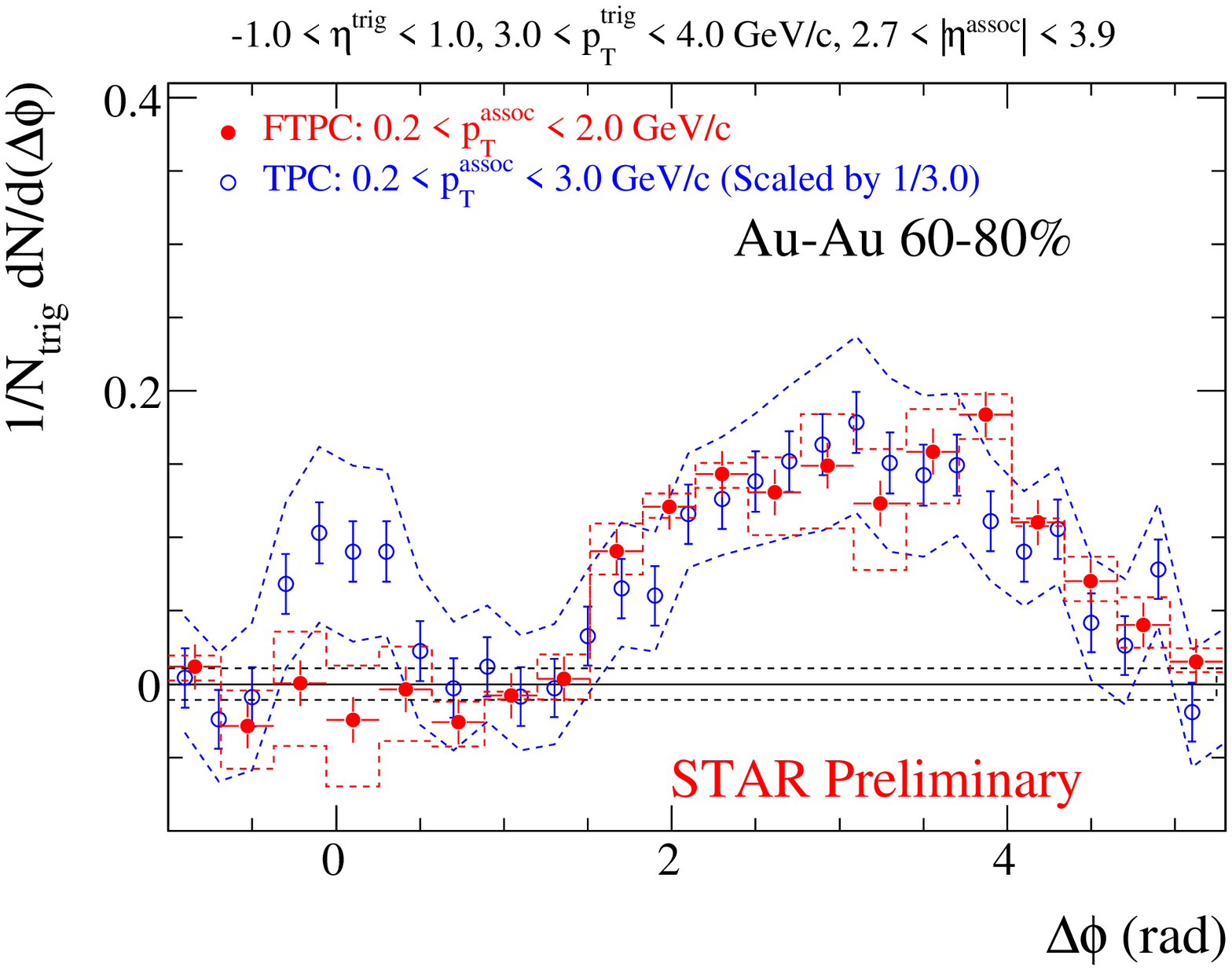}}
\resizebox{.327\textwidth}{!}{\includegraphics{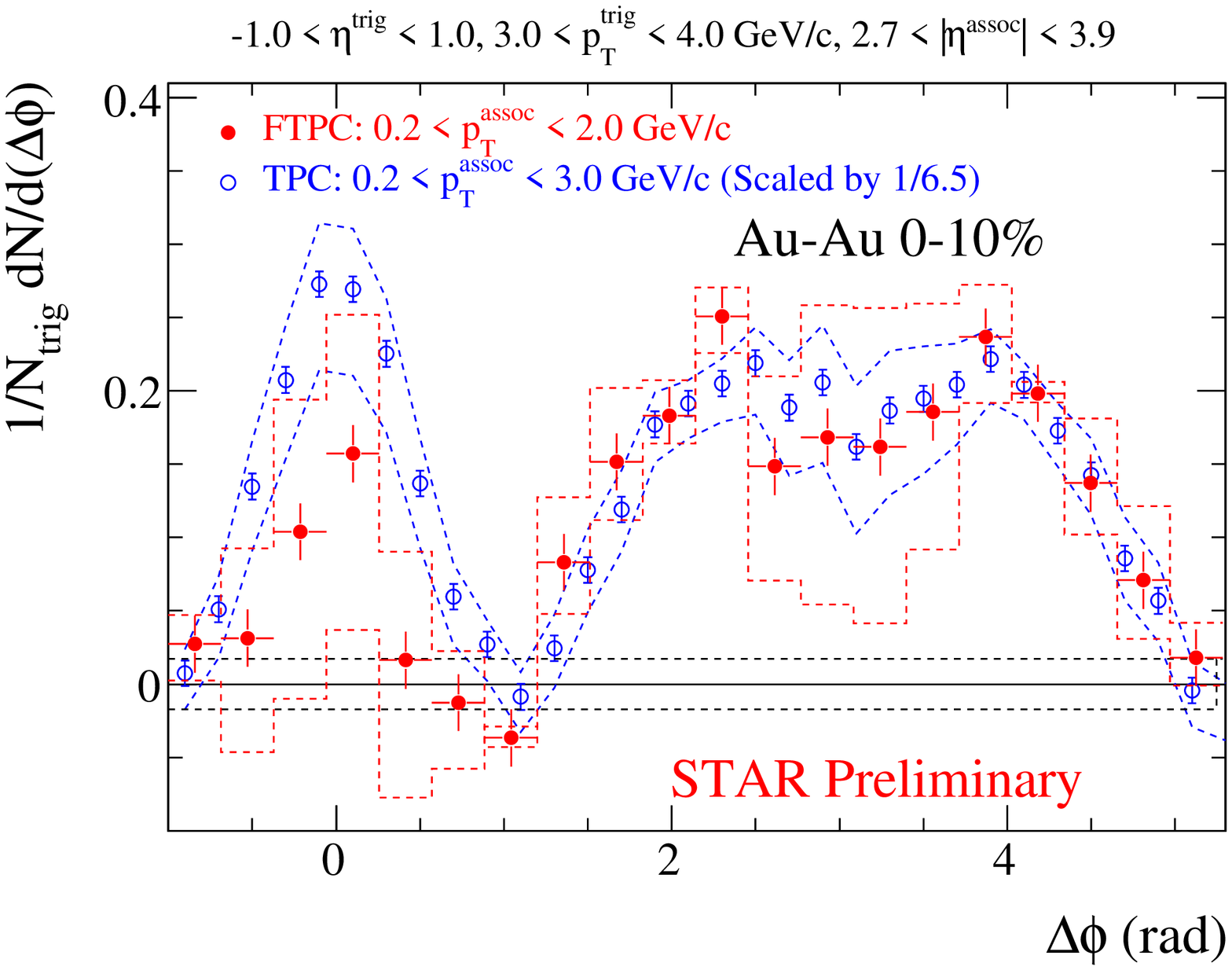}}
}
\caption{(color online) Azimuthal correlations at forward rapidities (red) compared to those at mid-rapidity (blue) in d+Au (left panel), 80-60\% peripheral Au+Au (middle panel), and top 10\% central Au+Au collisions (right panel). The trigger particles are within $3<\pt^{\rm trig}<4$~GeV/$c$ and $|\eta^{\rm trig}|<1$, and the associated particles are within $0.2<\pt< 2$~GeV/$c$ for forward rapidities $2.7<|\eta|<3.9$ and $0.2<\pt<3$~GeV/$c$ for mid-rapidity $|\eta|<1$. The mid-rapidity results are scaled to facilitate comparison of the correlation shapes. Dashed lines represent systematic uncertainties of the Au+Au data.}
\label{fig:comparison}
\end{figure}

Figure~\ref{fig:comparison} compares the shapes of the azimuthal correlations at mid- and forward rapidities. The near-side correlations are different because the near-side correlation centers around the trigger particle which is at mid-rapidity in the main TPC and should be naively absent from forward rapidities (except for possibly the ridge). The away-side correlations are remarkably similar between mid- and forward rapidities. One should note that the individual correlation functions have large systematic uncertainties, but these uncertainties (mainly due to flow) are strongly correlated. Therefore, while the particular shape is subject to a large uncertainty, the comparison between the shapes is rather robust. 
The broadening and flattening of the away-side shape indicates that partons at forward rapidity suffer strong interactions with the medium. The similarity of the away-side shapes at mid- and forward rapidity is rather striking, given the different multiplicity (medium density) and geometry (path length).
This could imply a number of scenarios: (1) the energy loss is so large that the partons lose almost all their energy and the different energy loss rates for gluons and quarks are insignificant; (2) the forward-rapidity partons (mostly quarks) traverse more matter (longitudinally) than the mid-rapidity partons (mostly gluons), almost perfectly cancelling the difference in their energy loss rates; and (3) the physics may just be the same between gluon- and quark-jets and between mid- and forward rapidities. Futher studies are needed in order to discriminate these different scenarios. Experimentally, for instance, Cu+Cu collisions and more peripheral Au+Au data may shed more light on (1) and moving to other forward rapidities may give insight into (2). Theoretically, calculations of parton energy loss and the resulting angular correlations at both mid- and forward-rapididity will be essential to disentangle the underlying physics processes.

\section{Conclusions}
\label{conclusion}

We have presented jet-like azimuthal correlations of charged hadrons at forward rapidities ($2.7<|\eta|<3.9$, $\pt<2$~GeV/$c$) measured by the STAR FTPCs with high-$\pt$ charged trigger particles at mid-rapidity ($|\eta|<1$, $\pt>3$~GeV/$c$) measured by the main TPC. Preliminary results are reported for $pp$, d+Au, and Au+Au collisions at $\snn=200$~GeV.

Near-side correlations are absent from $pp$ and d+Au collisions. 
Near-side correlations in Au+Au collisions are consistent with zero within the systematic uncertainties for $0.2<\pt<2$~GeV/$c$. However, the high associated $\pt$ data ($\pt>1$~GeV/$c$) in central Au+Au collisions suggest a finite near-side correlation, which may indicate the presence of long range $\deta$ correlations at forward rapidities.

Away-side jet-like correlation signals are observed at forward rapidities. While the correlation shapes are similar, the correlation strength at the d-side is a factor of 2 smaller than that at the Au-side in d+Au collisions, and the $pp$ result is approximately equal to the average of the two. The gluon anti-shadowing and the EMC effect would yield an enhanced d-side correlation.
The data are in qualitative agreement with small-$x$ gluon suppression in the Au-nucleus by the Color Glass Condensate. 
Quantitative calculations of the Color Glass Condensate and possibly other physics mechanisms, such as energy degrading of the incoming deuteron from multiple scatterings in the Au-nucleus, are needed in order to further our understanding.

The away-side correlation in Au+Au collisions at forward rapidities broadens with centrality. The correlation shapes are similar to those at mid-rapidity. The similarity suggests a similar energy loss picture at mid- and forward rapidities. The underlying physics mechanisms require further investigations.





\begin{thebibliography}{00}




\bibitem{whitepapers}
J. Adams {\it et al.} (STAR Collaboration), Nucl. Phys. {\bf A757} (2005) 102;
K. Adcox {\it et al.} (PHENIX Collaboration), Nucl. Phys. {\bf A757} (2005) 184.

\bibitem{FPD}
J. Adams {\it et al.} (STAR Collaboration), nucl-ex/0602011.

\bibitem{PHENIXmuon}
S.S. Adler {\it et al.} (PHENIX Collaboration), Phys. Rev. Lett. {\bf 96} (2006) 222301.

\bibitem{jetspectra}
J. Adams {\it et al.} (STAR Collaboration), Phys. Rev. Lett. {\bf 95} (2005) 152301.

\bibitem{otherpapers}
F. Wang (STAR Collaboration), J. Phys. Conf. Ser. {\bf 27} (2005) 32 [nucl-ex/0508021];
J.G. Ulery (STAR Collaboration), Nucl. Phys. {\bf A774} (2006) 581 [nucl-ex/0510055]; 
S.S Adler {\it et al.} (PHENIX Collaboration), Phys. Rev. Lett. {\bf 97} (2006) 052301;
M.J. Horner (STAR Collaboration), arXiv:nucl-ex/0701069.

\bibitem{UleryHP06}
J.G. Ulery, Nucl. Phys. {\bf A783} (2007) 511, arXiv:nucl-ex/0609047.
J.G. Ulery (STAR Collaboration), Quark Matter 2006 poster proceedings.

\bibitem{Puschke}
J. Putschke (STAR Collaboration), arXiv:nucl-ex/0701074.

\bibitem{pQCDeloss}
R. Baier, D. Schiff, B.G. Zakharov, Annu. Rev. Nucl. Part. Sci. {\bf 50} (2000) 37.


\bibitem{PDF}
H.L. Lai {\it et al.}, Eur. Phys. J. C {\bf 12} (2000) 375.

\bibitem{CGC}
L.V. Gribov, E.M. Levin and M.G. Ryskin, Phys. Rep. 100 (1983) 1;
D. Kharzeev, E. Levin and L. McLerran, Nucl. Phys. {\bf A748} (2005) 627.

\bibitem{TPC}
M. Anderson {\it et al.}, Nucl. Instrum. Meth. {\bf A499} (2003) 659.

\bibitem{FTPC}
K.H. Ackermann {\it et al.}, Nucl. Instrum. Meth. {\bf A499} (2003), 713.

\bibitem{v2MRP}
J. Adams {\it et al.} (STAR Collaboration), Phys. Rev. C {\bf 72} (2005) 014904.


\bibitem{WangHP06}
F. Wang (STAR Collaboration), Nucl. Phys. {\bf A783} (2007) 157, arXiv:nucl-ex/0610011.
L. Molnar (STAR Collaboration), arXiv:nucl-ex/0701061.

\bibitem{brahms}
I. Arsene {\it et al.} (BRAHMS Collaboration), Phys. Rev. Lett. {\bf 93} (2004) 242303.

\bibitem{EKS98}
K.J. Eskola, V.J. Kolhinen and P.V. Russkanen, Nucl. Phys. B {\bf 535} (1998) 351.

\bibitem{EMC}
J.J. Aubert {\it et al.}, Phys. Lett. B {\bf 123} (1983) 275;
J. Gomez {\it et al.}, Phys. Rev. D {\bf 49} (1994) 4348.

\bibitem{method}
J.G. Ulery and F. Wang, arXiv:nucl-ex/0609016.

\end{thebibliography}
\end{document}